\documentclass[aps,twocolumn]{revtex4}
\usepackage{dcolumn}
\usepackage{graphicx}
\usepackage{amsmath}
\usepackage{amsfonts}
\usepackage{amssymb}
\usepackage{psfrag}
\usepackage{wrapfig}
\usepackage{subfigure}
\usepackage{makeidx}
\usepackage{bm}
\usepackage{epsf}

\begin{document}

\title{ Localized Nonlinear Waves in a Two-Mode Nonlinear Fiber}
\author{Li-Chen Zhao$^{1,2}$}
\author{Jie Liu$^{1,3}$}\email{liu_jie@iapcm.ac.cn}
\address{$^1$Science and Technology Computation Physics Laboratory,
 Institute of Applied Physics and Computational Mathematics, Beijing 100088,
China}
\address{$^2$Department of Modern Physics, University of Science and Technology of China, Hefei 230026, China}
\address{$^3$Center for Applied Physics and Technology, Peking University, Beijing 100084,
China}
\date{\today}
\begin{abstract}
We find that diverse nonlinear waves such as soliton, Akhmediev
breather, and rogue waves, can emerge and interplay with each other
in a two-mode coupled system. It provides a good platform to study
interaction between different kinds of nonlinear waves. In
particular, we obtain dark rogue waves analytically for the first
time in the coupled system, and find that two rogue waves can appear
in the temporal-spatial distribution. Possible ways to observe these
nonlinear waves are discussed.

\end{abstract}

 \maketitle

\section{Introduction}
Solitons are localized waves arising from the interplay between
self-focusing (self-defocusing) and dispersion effect, and could be
one of the intense studies in nonlinear science. The well-known
solitons of the scalar nonlinear Schr$\ddot{o}$dinger equation
(NLSE) have been studied in many different systems including
plasmas, optical fibers and cold atoms, mainly including bright
soltons \cite{1,2,3,4} and dark solitons \cite{Burger,6,7,8,9,Wu}.
In addition to solitons, the NLSE admits other classes of localized
structure, such as Akhmediev breather (AB) \cite{N.Akhmediev}, and
rogue waves (RW) \cite{Shrira}. Moreover, the AB and RW have been
observed in one-mode nonlinear fibers experimentally \cite{Kibler}.

Recently, the studies for solitons have been extended to
multi-component coupled systems. It has been reported that the
bright-bright(B-B), dark-dark(D-D), bright-dark(B-D) solitons can
exist in different parameters regimes
\cite{Kockaert,Park,Kanna,Lakshmanan,Zhao}. The bright-dark soliton
dynamics have been observed experimentally in \cite{Becker}. Besides
these solitons, studies have indicated that there exist other kinds
of nonlinear localized waves in the coupled system, such as AB
\cite{Park,Forest}, and vector RWs \cite{Ling2}. These studies
indicate that nonlinear waves in coupled system are much more
diverse than the ones in uncoupled systems.

In this paper, we study diverse families of nonlinear localized
waves in a two-mode optical fiber. The proper conditions for these
nonlinear waves are given explicitly. Notably, interactions between
different kinds of nonlinear waves can be observed analytically. For
example,  bright or dark soliton interplay with RW could be
observed, which provides a good platform to study interaction
between RW and other nonlinear waves. Moreover, we prove that dark
RW indeed can exist in one component of the coupled system, and
report that two RWs could appear in temporal-spatial distribution.
Especially, there are some additional requirements on the nonzero
backgrounds for vector RWs, which is quite different from the RW in
single-component nonlinear system. As an example to demonstrate
possibilities to observe these nonlinear waves, we present one
possible way to create RW in the two-mode nonlinear fiber system.

The paper is organized as follows. In Section {\rm II}, we present
 three families of nonlinear localized wave solutions and discuss
the explicit conditions under which they could exist in detail. In
Section {\rm III}, the possibilities to observe them are discussed.
The conclusion and discussion are made in Section {\rm IV}.

\section{ Analytical vector nonlinear wave solutions}
It is well known that coupled NLS equations are often used to
describe the interaction among the modes in nonlinear optics. We
begin with the well known two-coupled NLSE in dimensionless form
\begin{eqnarray}
 &&i\frac{\partial \psi_1}{\partial z}+\sigma_1 \frac{\partial^2
\psi_1}{\partial t^2}+[2g_1|\psi_1|^2+2g_2|\psi_2|^2]\psi_1 =0,\\
&&i\frac{\partial \psi_2}{\partial z}+\sigma_2\frac{\partial^2
\psi_2}{\partial t^2}+[2 g_1|\psi_1|^2+2g_2|\psi_2|^2]\psi_2=0,
\end{eqnarray}
where $\psi_1$ and $\psi_2$ are complex envelopes of the electric
fields of the two modes.  $z$ denotes propagation distance, and $t$
represents the retarded time. The $\sigma_1$ and $\sigma_2$ depend
on the signs of the group velocity dispersion(GVD) in each mode,
i.e., $\sigma = +1$ for anomalous GVD and $\sigma = -1$ for normal
GVD. Here, we just consider the anomalous GVD case,
$\sigma_1=\sigma_2=1$. $g_1, g_2$ are nonlinear parameters
determined by properties of the Kerr medium with electrostrictive
mechanism of nonlinearity \cite{Afanasyev}. When $g_1=g_2=g$, it
will become the well-known Manakov model \cite{Park, Haelterman}.

The Eq.(1) has been solved to get soliton solution on trivial
background through Hirota bilinear method in \cite{Lakshmanan}.
Performing Darboux-transformation from a trivial seed solution, one
could get the bright-bright solitons \cite{Zhao}. It has been
reported that solitons could collide inelastically and there are
shape-changing collisions for coupled system, which are different
from uncoupled system \cite{Lakshmanan}. However, it is not possible
to study B-D soliton, ABs or RWs on trivial background. Next, we
will solve it from nontrivial seed solutions. The nontrivial seed
solutions are derived as follows
\begin{eqnarray}
\psi_{10}&=&s_1 \exp{[i \theta_1[t,z]]}\nonumber\\
&=&s_1 \exp{[ik_1t+i (2g_1 s_1^2+2g_2
s_2^2-i k_1^2) z]},\\
\psi_{20}&=&s_2 \exp{[i \theta_2[t,z]]}\nonumber\\
&=&s_2 \exp{[ik_2 t+i (2g_1 s_1^2+2g_2s_2^2-i k_2^2) z]},
\end{eqnarray}
where $s_1$ and $s_2$ are two arbitrary real constants, and denote
the backgrounds in which localized nonlinear waves emerge. $k_1$ and
$k_2$ denote the frequencies of the plane wave background in the two
modes respectively. We will solve the coupled system analytically
through Darboux transformation method, which can transfer nonlinear
problem to linear one. The corresponding Lax-pair of Eq.(1) with
$g_1, g_2>0$ can be derived as
\begin{eqnarray}
\partial_t\left(
\begin{array}{c}
\Phi_{1} \\ \Phi_{2 }\\\Phi_3\end{array}
 \right)
&=&U \left(
\begin{array}{c}
\Phi_{1} \\ \Phi_{2}\\ \Phi_3\end{array}
 \right),\\
\partial_z\left(
\begin{array}{c}
\Phi_{1} \\ \Phi_{2}\\\Phi_3\end{array}
 \right)&=&V\left(
\begin{array}{c}
\Phi_{1} \\ \Phi_{2}\\\Phi_3\end{array}
 \right),
\end{eqnarray}
where
\begin{equation*}
 U=\left(
\begin{array}{ccc}
-i \frac{2}{3}\lambda&\sqrt{g_1}\psi_1&\sqrt{g_2}\psi_2\\
-\sqrt{g_1} \bar{\psi}_1&\frac{i}{3}\lambda&0\\
-\sqrt{g_2} \bar{\psi}_2&0&\frac{i}{3}\lambda
\end{array}
 \right),
\end{equation*}
\begin{equation*}
 V=U\lambda+\left(
\begin{array}{ccc}
J_1& i\sqrt{g_1}\psi_{1t} &i\sqrt{g_2}\psi_{2t}\\
i\sqrt{g_1} \bar{\psi}_{1t}& J_2&-i\sqrt{g_1 g_2}
\psi_2\bar{\psi}_1\\i\sqrt{g_2} \bar{\psi}_{2t}&-i\sqrt{g_1 g_2}
\bar{\psi}_2\psi_1&J_3
\end{array}
 \right),
 \end{equation*}
and
 \begin{eqnarray}
J_1&=&ig_1 |\psi_1|^2+ig_2|\psi_2|^2\nonumber\\
J_2&=&-ig_1 |\psi_1|^2,\nonumber\\
J_3&=&-ig_2 |\psi_2|^2.\nonumber
\end{eqnarray}
Hereafter, the overbar denotes the complex conjugate. With spectral
parameter $\lambda$ and the nontrivial seed solutions, one can solve
the following equation of $\tau$ to get the diagonalized or Jordan
forms of $U$ and $V$ through the method presented in \cite{Park}.
Then, one could get nonlinear wave solutions through performing
Darboux transformation method. The equation of $\tau$, which is the
eigenvalue equation of the transformed $U$, is calculated as
\begin{eqnarray}
 \tau^3+ a_1 \tau +b_1=0,
 \end{eqnarray}

where \begin{eqnarray}
 a_1&=&b c-a b-a c+g_1 s_1^2 +g_2
s_2^2,\nonumber\\
b_1&=&a b c-g_1 s_1^2 c-g_2 s_2^2 b,\nonumber\\
 a&=&i 2
\frac{\lambda }{3}+\frac{i }{3}
(k_1+k_2),\nonumber\\
b&=&i \frac{\lambda }{3}+\frac{i}{3} (2
k_1-k_2),\nonumber\\
c&=&i \frac{\lambda }{3}+\frac{i}{3} (2 k_2-k_1).\nonumber
\end{eqnarray}

There will be three cases for the roots $\tau_j$. Each one
corresponds to a different family of nonlinear waves in the coupled
systems.

Case1: The three roots are all different, we find that there are
bright-dark solitons pair interplay with Akhmediev breather (B-DAB),
and Akhmediev breathers-Akhmediev breathers (ABs-ABs) solutions in
the coupled system.

Case2: There are one single root and double roots, the bright-dark
interplay with rogue waves (B-DRW), Akhmediev breather-Akhmediev
breather interplay with RW (AB-ABRW) solutions could exist in the
coupled system.

Case3: There are triple roots, the RWs with no other type nonlinear
waves solutions could exist with certain conditions.

We find that there are many explicit requirements on signals or
backgrounds for different kinds of vector nonlinear waves. It should
be pointed out that similar studies have been done on nontrivial
backgrounds in \cite{Park, Forest,Ling2,Baronio}. Distinct from
them, we investigate all possible cases and report explicit
conditions for these nonlinear wave solutions in the nonlinear
coupled systems. The whole analysis would help us to understand the
relations between vector solitons, breathers, and RWs. Furthermore,
we report that there are some new types of nonlinear waves, such as
dark rouge waves, just two rouge waves in temporal-spatial
distribution. Next, we will discuss these three circumstances.

\subsection{Bright-dark solitons interplay
 with Akhmediev breather, and Akhmediev breathers-Akhmediev breathers}
   When $\tau_j$ (j=1,2,3) are three different roots for Eq.(7), we can
solve the Lax-pair to get
 $\Phi_1$, $\Phi_2$, $\Phi_3$ as
\begin{eqnarray}
\Phi_1[t,z]&=&(\Phi_{01}+ \Phi_{02}+\Phi_{03})\nonumber\\
&&\times \text{Exp}\left[\frac{I}{3} (\theta_1[t,z]+\theta
_2[t,z])\right],\\
\Phi_2[t,z]&=&\left(-\frac{\sqrt{g_1} s_1}{\tau_1-b}
\Phi_{01}-\frac{\sqrt{g_1}s_1}{\tau_2-b} \Phi _{02}-\frac{\sqrt{g_1}
s_1}{\tau_3-b}\Phi_{03}\right)\nonumber\\
&&\times \text{Exp}\left[\frac{I}{3}
(\theta_2[t,z]-2 \theta_1[t,z])\right],\\
\Phi_3[t,z]&=&\left(-\frac{\sqrt{g_2}s_2}{\tau_1-c}
\Phi_{01}-\frac{\sqrt{g_2}s_2}{\tau_2-c} \Phi
_{02}-\frac{\sqrt{g_2}s_2}{\tau_3-c} \Phi_{03}\right)\nonumber\\
&&\times \text{Exp}\left[\frac{I}{3} (\theta_1[t,z]-2 \theta
_2[t,z])\right],
 \end{eqnarray}
 where
 \begin{eqnarray}
\Phi_{01}[t,z]&=&A_1 \text{Exp}[\tau_1 t+I \tau_1^2 z+2(\lambda
-k_1-k_2) \tau_1 z/3]\nonumber\\
&&\times \text{Exp}[ F[z]],\nonumber\\
\Phi_{02}[t,z]&=&A_2 \text{Exp}[\tau_2 t+I \tau_2^2 z+2(\lambda
-k_1-k_2) \tau_2 z/3]\nonumber\\
&&\times \text{Exp}[ F[z]],\nonumber\\
\Phi_{03}[t,z]&=&A_3 \text{Exp}[\tau_3 t+I \tau_3^2 z+2(\lambda
-k_1-k_2) \tau_3 z/3]\nonumber\\
&&\times \text{Exp}[ F[z]],\nonumber
\end{eqnarray}
where $\lambda=a_0+ib_0$ and $F[z]=2 I/9(\lambda^2 +(k_1+k_2)\lambda
+k_1^2+k_2^2-k_1 k_2+3g_1 s_1^2+3g_2 s_2^2)z$. Between these
expressions, the parameters $a_0$, $ b_0$, $A_j(j=1,2,3)$, and
$s_1$, $s_2$ are all real numbers which relate with the initial
condition of soliton or AB, such as initial coordinate , initial
velocity, and initial shape. Then, one can perform the following
Darboux transformation to get one family of solutions for Eq.(1) and
(2)
\begin{eqnarray}
\psi_1&=&\psi_{10}-\frac{1}{\sqrt{g_1}}\frac{i(\lambda-\bar{\lambda})\bar{u}}{1
+|u|^2+|v|^2},\\
\psi_2&=&\psi_{20}-\frac{1}{\sqrt{g_2}}\frac{i(\lambda-\bar{\lambda})\bar{v}}{1
+|u|^2+|v|^2},
\end{eqnarray}
where $u=\frac{\Phi_2}{\Phi_1}$ and $v=\frac{\Phi_3}{\Phi_1}$.
 This is a generic
solution which can be used to study the properties of B-D solitons,
AB-AB, B-AB, and D-AB for coupled nonlinear systems. To get B-B
solitons, it is much easier to solve the Lax-pair form trivial seed
solution, namely, $\psi_{10}=\psi_{20}=0$. Based on the generic
solution, it is more convenient to study their dynamics on different
nontrivial backgrounds for the coupled systems. We find that there
are much more abundant nonlinear waves on the nontrivial background
than the ones on the trivial background.

\begin{figure}[htb]
\centering
\subfigure[]{\includegraphics[height=30mm,width=40mm]{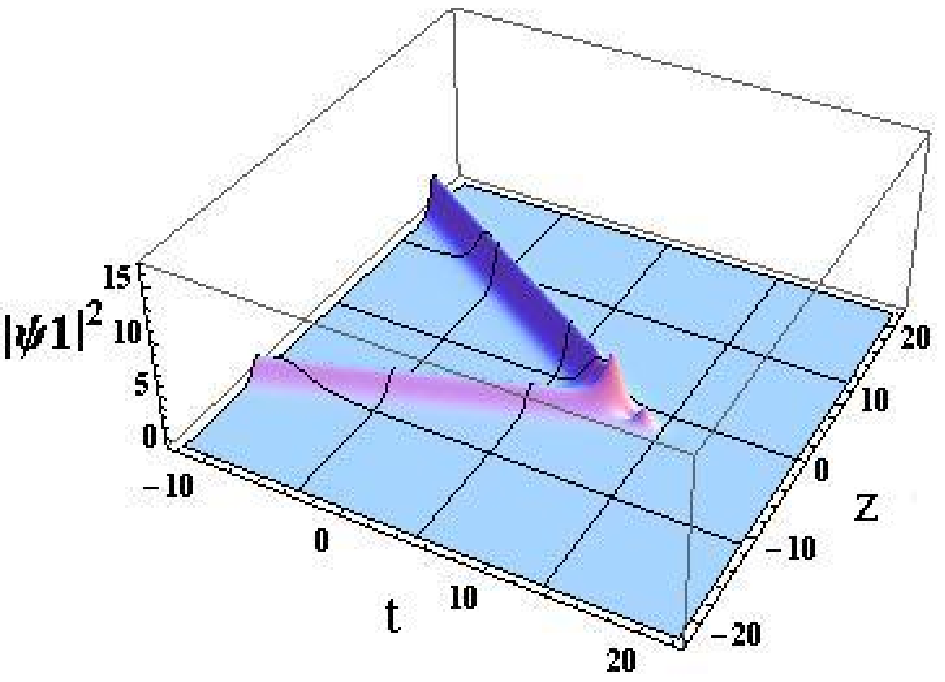}}
\hfil
\subfigure[]{\includegraphics[height=30mm,width=40mm]{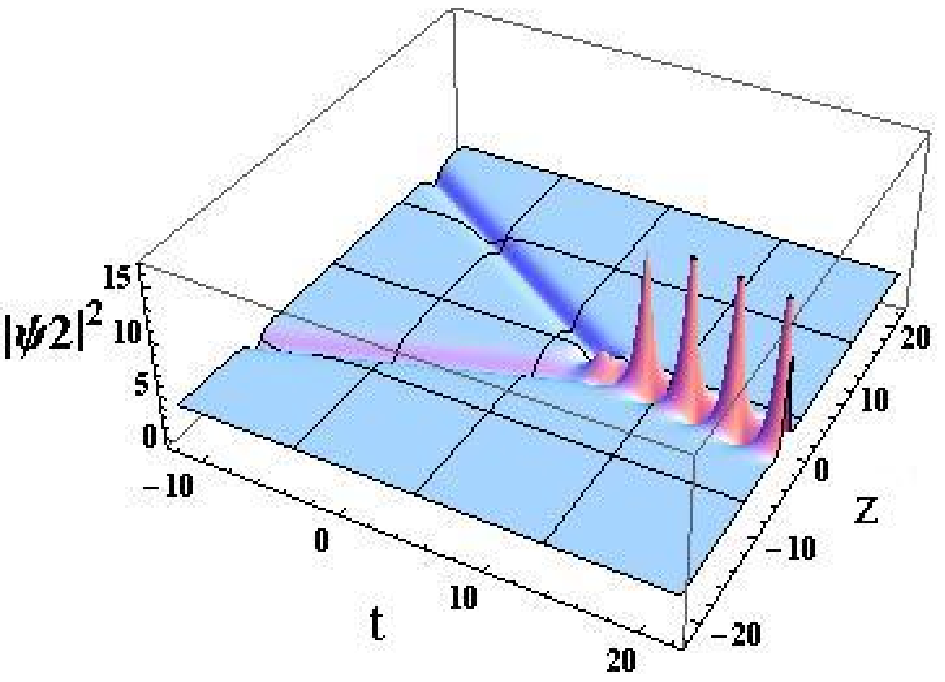}}
\caption{(color online) The evolution plot of B-DAB in the coupled
system, (a) for bright soliton in $\psi_1$ component and (b) for
dark with an AB in $\psi_2$ component. It is seen that the bright
soliton in one field is reflected like hard wall by the AB in the
other one. The dark soliton is reflected well by the AB too. The
coefficients are $a_0 = 0, b_0 = 1.2, g_1= g_2 = 0.25, s_1 = 0.001,
s_2 = 2, A_1 = 1, A_2 = 4, A_3 = 3, k_1 = -0.5$, and $ k_2 = 0.5$.}
\end{figure}

When $s_1\rightarrow 0$, $s_2\neq 0$, with the certain parameters,
the bight-dark interplay with AB (B-DAB) could exist, such as Fig.
1. Particularly, Fig. 1(a) shows that one bright soliton in $\psi_1$
component is reflected by the other one's density distribution.
Considering that the density distribution of one component could be
seen as nonlinear potential distribution for solitons in the other
field, it is not hard to understand the refection effect.
Correspondingly, one can observe a dark soliton collide with a AB
and its shape is changed in Fig. 1(b). The reflecting inelastic
collisions happen between nonlinear waves with different structures.
Considering that Becker et al. have observed dark soliton collide
with the B-D soliton in BEC \cite{Becker}, it is believed there are
some possibilities to observe the refection collision presented
here.
\begin{figure}[htb]
\centering
\subfigure[]{\includegraphics[height=30mm,width=40mm]{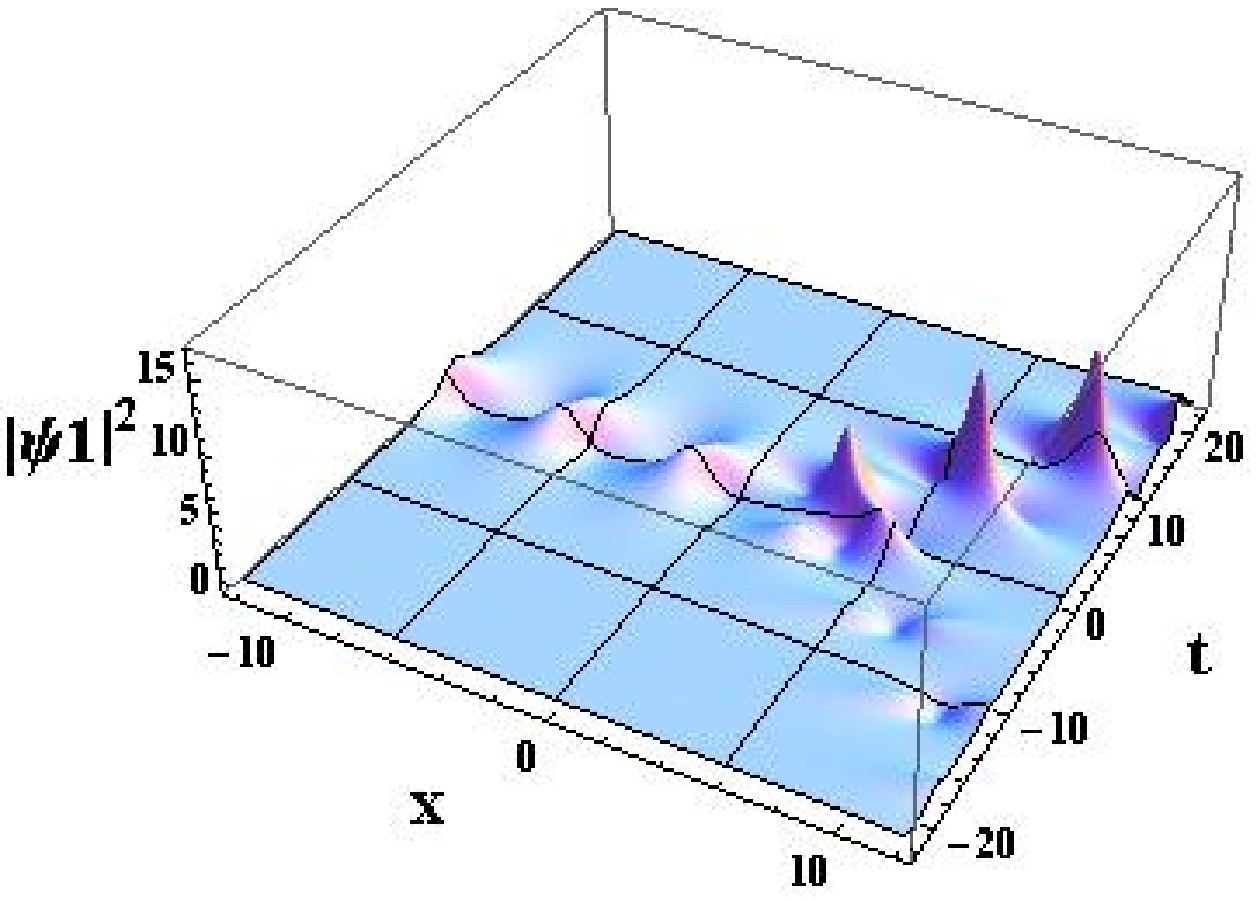}}
\hfil
\subfigure[]{\includegraphics[height=30mm,width=40mm]{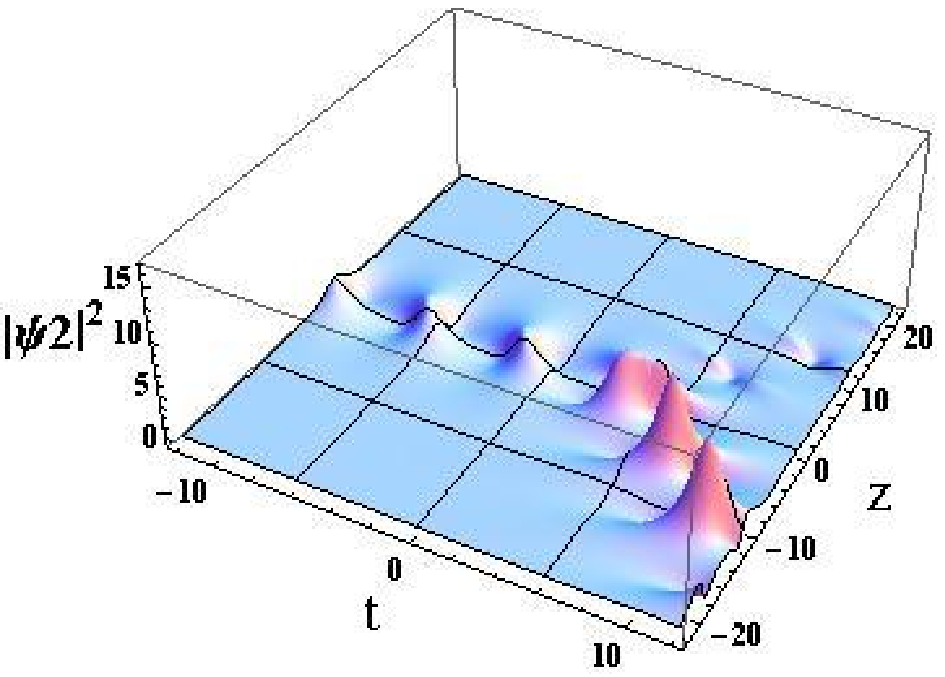}}
\caption{(color online)   The evolution plot of ABs-ABs in two
components coupled system. (a) for ABs soliton in $\psi_1$ component
and (b) for ABs in $\psi_2$ component. It is shown that one AB
collides with a AB inelastically. The coefficients are $a_0 = 0, b_0
= 1.2, g_1 =g_2 = 0.25, s_1 =1, s_2 = 1, A_1 = 1, A_2 = 4, A_3 = 3,
k_1 = -0.5$, and $ k_2 = 0.5$.}
\end{figure}

When $s_1\neq 0$ and $s_2 \neq 0$, there are many kinds of different
soliton and AB appearing in the coupled system. In the nontrivial
background, ABs-ABs could exist. The reflection between AB and AB
can be observed, just as shown in Fig. 2(a). Two ABs could collide
inelastically and then merge into one AB, as shown in Fig. 2(b).
Inelastic collision for dark, bright, AB with AB could exist under
some certain conditions. These characters could be used to design
optical switches in nonlinear optics.

\subsection{Bright-dark solitons or Akhmediev Breather interplay
 with Rogue Wave}
It is found that when the equation of $\tau$ has double roots,
namely, $\tau_1=-2\tau_2$ and $\tau_2=\tau_3$, there is a certain
requirement on the spectral parameter $\lambda$. The parameter
$\lambda$ which mainly determines the form of initial signals should
satisfy the equation as follows
\begin{eqnarray}
&&[3(k_1+k_2)^2+4(C-A)]
\lambda^4\nonumber\\
&&+[4(k_1+k_2)^3+(k_1+k_2)(8C-6A)-4B]\lambda^3\nonumber\\
&&+[4(k_1+k_2)^2C+4C^2/3-A^2-6(k_1+k_2)B]\lambda^2\nonumber\\
&&+[4(k_1+k_2)C^2/3-2 A B]\lambda +\frac{4C^3}{27}-B^2=0,
\end{eqnarray}
where
\begin{eqnarray}
A&=&2(2k_1-k_2)(2k_2-k_1)+(k_1+k_2)^2+9g_1 s_1^2+9 g_2
s_2^2,\nonumber\\
B&=& 9g_1s_1^2(2k_2-k_1)+(k_1+k_2)(2k_1-k_2)(2k_2-k_1)\nonumber\\
&&+9g_2 s_2^2(2k_1-k_2),\nonumber\\
C&=&(k_1+k_2)^2-(2k_1-k_2)(2k_2-k_1)+9g_1 s_1^2+9 g_2
s_2^2.\nonumber
\end{eqnarray}
 Under
this requirement condition, the roots of $\tau$ will be given as
\begin{eqnarray}
\tau_1&=&-2\tau_2,\\
\tau_2&=&\tau_3=\frac{H_1(\lambda)}{H_2(\lambda)},
\end{eqnarray}
 where
\begin{eqnarray}
H_1(\lambda)&=& I [2 \lambda^3+3(k_1+k_2) \lambda^2+(2 (2
k_1-k_2)(2k_2-k_1)\nonumber\\
&&+(k_1+k_2)^2+9 g_1 s_1^2+9g_2 s_2^2)\lambda \nonumber\\
&& +9 g_2 s_2^2 (2k_1-k_2)+9g_1 s_1^2 (2
k_2-k_1)\nonumber\\
&&+(k_1+k_2)(2 k_1-k_2)(2
k_2- k_1)],\nonumber\\
H_2(\lambda)&=& 6 \lambda ^2+6(k_1+k_2) \lambda +2( k_1+
k_2)^2\nonumber\\
&&-2(2 k_1- k_2)(2 k_2- k_1)+18  g_1 s_1^2+18 g_2 s_2^2.\nonumber
\end{eqnarray}
\begin{figure}[htb]
\centering
\subfigure[]{\includegraphics[height=30mm,width=40mm]{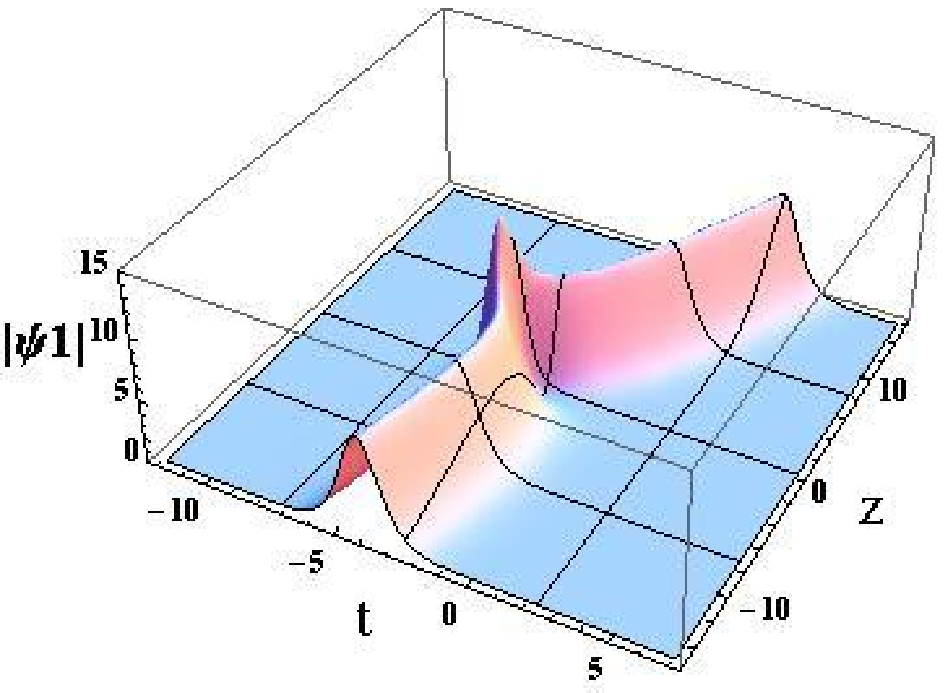}}
\hfil
\subfigure[]{\includegraphics[height=30mm,width=40mm]{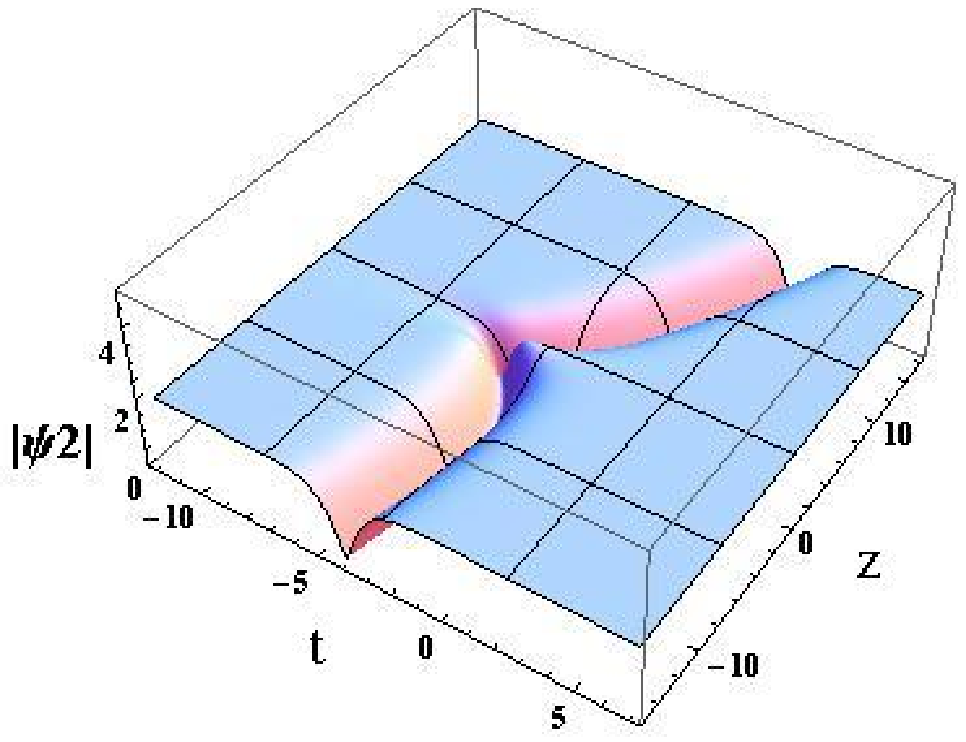}}
\caption{(color online)   The evolution of B-DRW in the coupled
system, (a) for one bright soliton with RW in $\psi_1$ component and
(b) for one dark with a RW in $\psi_2$ component. It is shown that
one dark soliton collides with RW elastically. Bright soliton are
attracted by RW when being next to it. The coefficients are $a_0 =
-0.1, b_0 = - 2, g_1 = g_2 = 0.25, s_1 =0.01, s_2 = 2, A_1 = 1, A_2
= 4, A_3 = 3, k_1 = -1$, and $ k_2 = 0.1$.}
\end{figure}

Under these conditions, one RW appears, namely, the AB in previous
part will evolve as a RW. And the RW with other type nonlinear waves
solution is given in Appendix A. One can see that the form of
solution partly includes rational functions. When the amplitude of
one component's background $s_1\rightarrow 0$, and $s_2>0$, the
bright-dark interplay with RW (B-DRW) could be observed, such as
Fig. 3. It is seen from Fig. 3(a) that the bright soliton is
attracted by the RW, and collide with it, then the shape of dark
soliton is unchanged after RW disappearing. When $s_1, s_2>0$, the
ABs emerge with RW (AB-ABRW) in each component, such as Fig. 4. One
normal RW appears around $t=0$ place near $z=0$ in the first
component field, and correspondingly, one dark RW emerges in the
second component, which can be well shown  by density distribution
plot near $z=0$. Moreover, the relative location between bright or
dark solitons and RW can be varied through changing the parameters
$A_j$. Especially, when $A_1=0$, the AB-ABRW solution will
correspond to pure one vector RW solution, shown in Fig. 4(c) and
(d). In Fig. 4(d), it is shown well that the dark RW indeed exists
in the second component. The similar dark RWs have been expected in
mixed BEC system through numeric stimulation in \cite{Bludov2}.
Namely, we could obtain analytical dark RW solution with the
corresponding parameters in Fig. 4(d). Furthermore, when $A_3=0$,
the solution will become AB-AB solution with no RW any more. The
parameter $A_2$ just affects the place where RW appears.

\begin{figure}[htb]
\centering
\subfigure[]{\includegraphics[height=30mm,width=40mm]{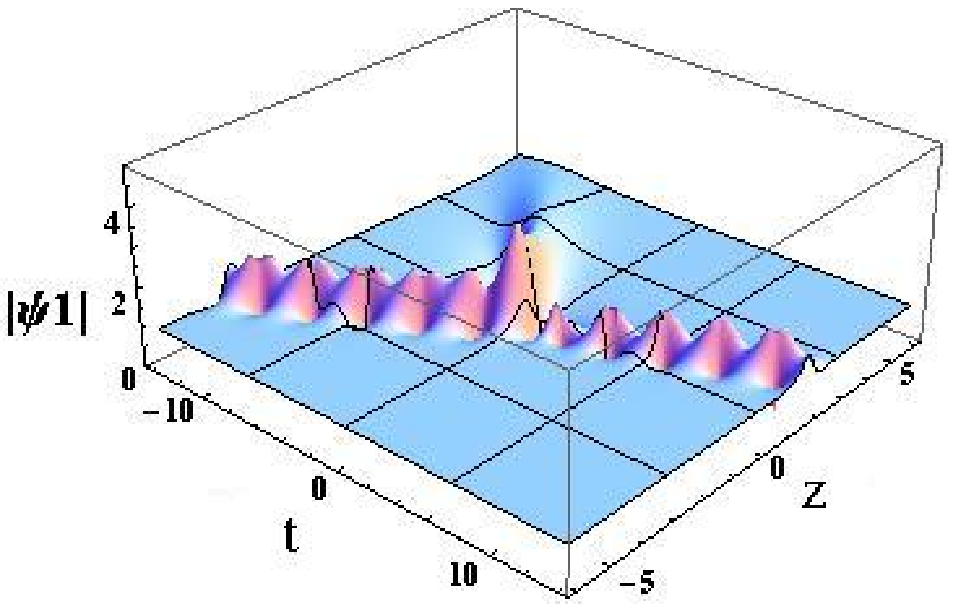}}
\hfil
\subfigure[]{\includegraphics[height=30mm,width=40mm]{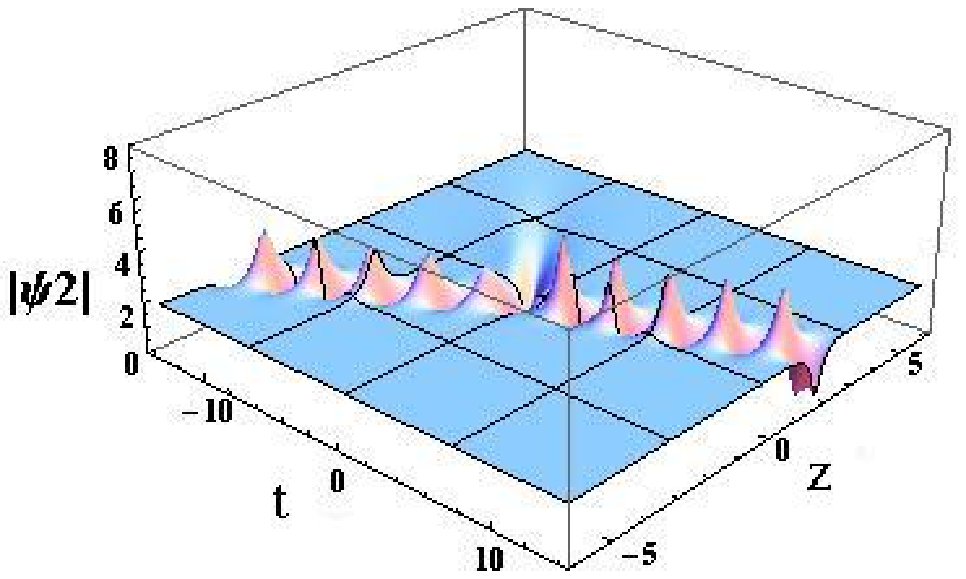}}
\hfil
\subfigure[]{\includegraphics[height=30mm,width=40mm]{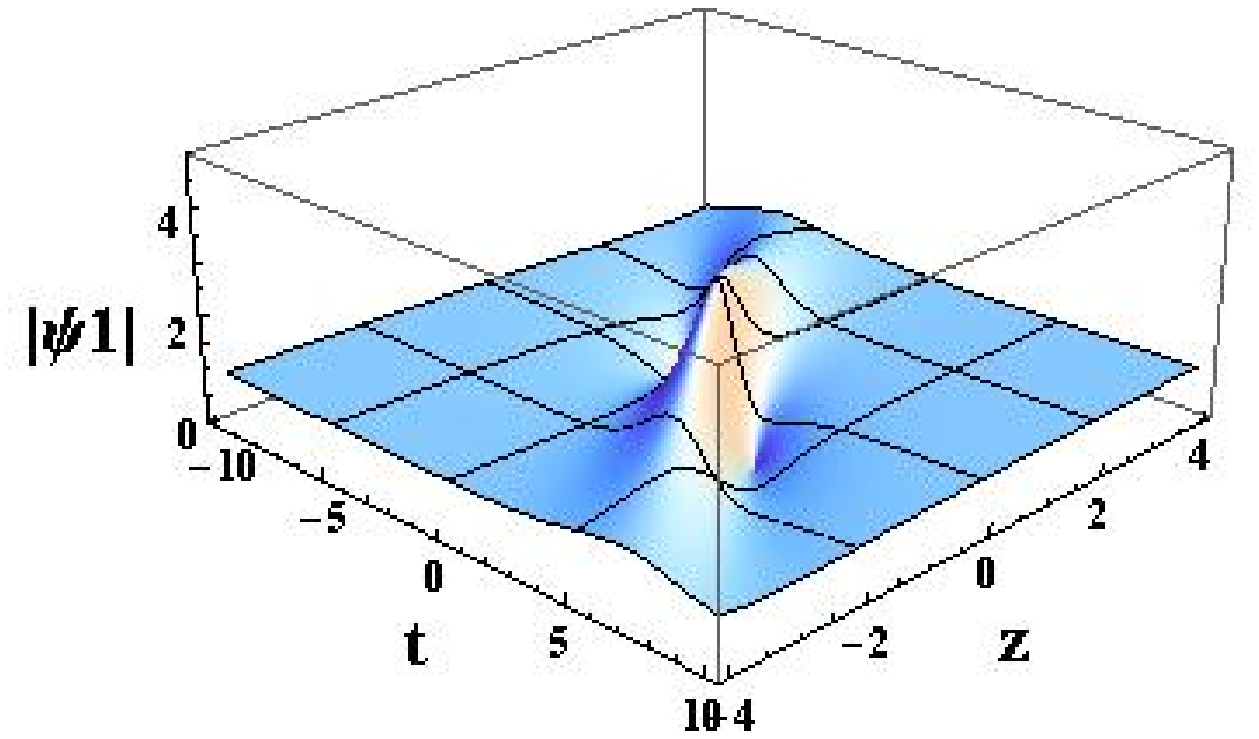}}
\hfil
\subfigure[]{\includegraphics[height=30mm,width=40mm]{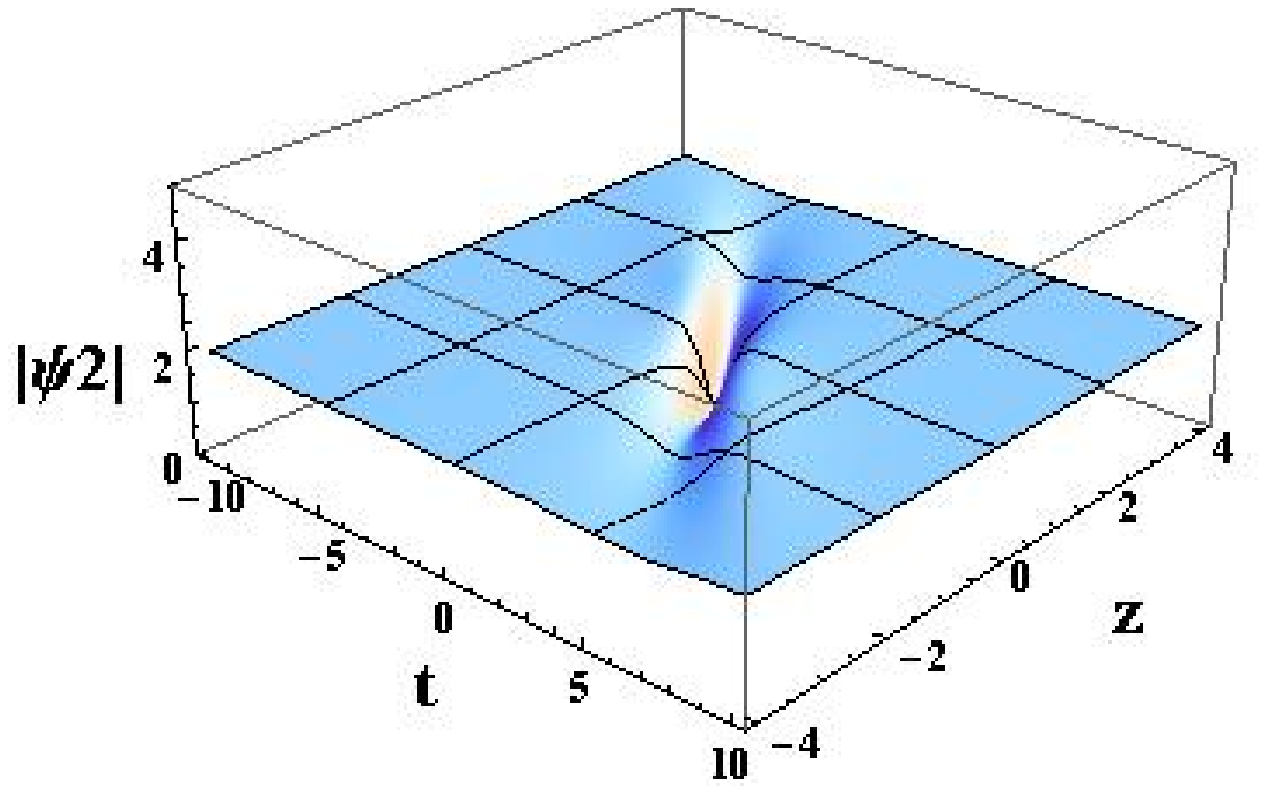}}
\caption{(color online)   The plot of AB-ABRW in the coupled system,
(a) for AB with a RW in $\psi_1$ component and (b) for AB with a
dark RW in $\psi_2$ component. For (a) and (b), $A_1 = 1, A_2 = 4$.
(c) The plot of one RW in $\psi_1$ component and (d) The plot of one
RW in $\psi_2$ component. It is seen that the RW in $\psi_2$
component is a dark one. (c) and (d) with $A_1 = A_2=0$. The other
coefficients are $a_0 = -0.53, b_0 =2.26, g_1 = g_2 = 0.5, s_1 =1,
s_2 = 2, A_3 = 3, k_1 = -1$, and $ k_2 = 0.1$.}
\end{figure}

 On arbitrary
nontrivial background, when the initial condition can be given as
nearly with the certain $\lambda$ before, there are many
possibilities to observe one RW interacted with bright, dark, or
ABs. These interaction phenomena are impossible to be observed for
scalar NLS system, for bright, dark solitons and RW can not exist
simultaneously with the same conditions. The generalized nonlinear
waves solution provides us a good platform to study the interactions
between RW and other type nonlinear waves.  On the other hand, the
nonlinear wave solution partly including rational functions
corresponds to a RW with other type nonlinear waves. When $A_1=0$,
the AB-ABRW solution will be all rational form and it is
 one vector RW solution. One could guess that the vector RWs with no other type nonlinear waves solution
  could have the all rational form, just like RW solution in uncoupled system.
Then, we will try to derive the all rational form solutions.

\subsection{ Rogue Waves with no other nonlinear waves}
When $\tau$ has triple roots, we can derive the solutions with all
rational functions, which usually correspond to RWs phenomena
\cite{Shrira,N.Akhmediev}. Interestingly, we find that there are
some certain requirements on the nontrivial backgrounds of the two
components for these RWs solutions. The conditions are required as
\begin{eqnarray}
g_1 s_1^2&=&g_2 s_2^2,\\
 |k_1-k_2|&=&\sqrt{g_1}s_1,
\end{eqnarray}
\begin{figure}[htb]
\centering
\subfigure[]{\includegraphics[height=30mm,width=40mm]{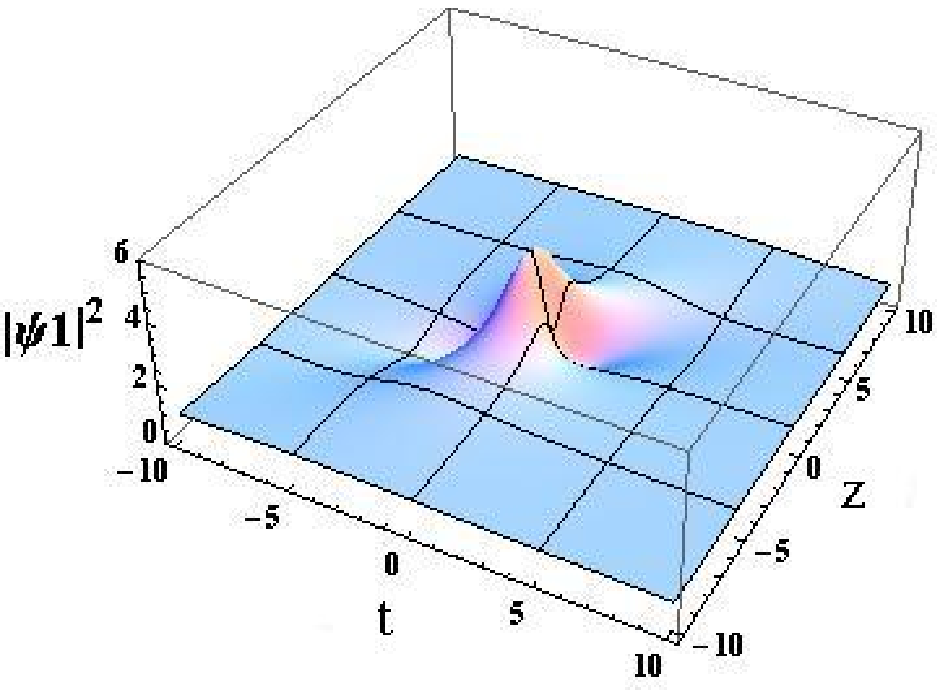}}
\hfil
\subfigure[]{\includegraphics[height=30mm,width=40mm]{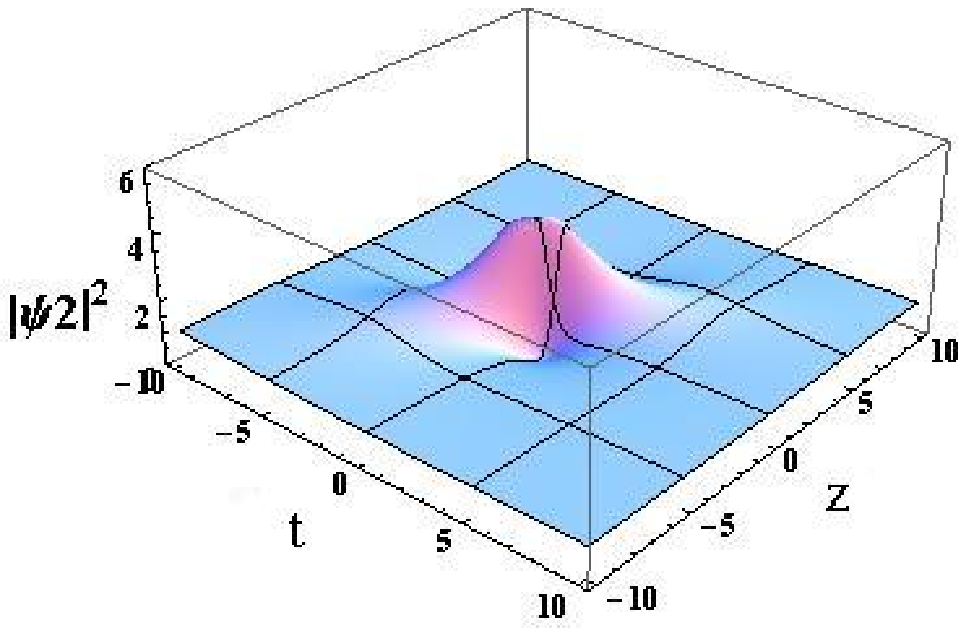}}
\caption{(color online)  The evolution plot of only one RW in
coupled system, (a) for one RW in $\psi_1$ component, (b) for one RW
in $\psi_2$ component. The coefficients are $g_1  = g_2 = 0.25, s_1
=s_2=1, A_1 = 1, A_2 = 1, A_3 = 0, k_1=0.55$, and $ k_2 = 0.05$.}
\end{figure}
which means that there are requirement on the nonlinear parameter
and the amplitude of each component, and the difference of their
wave vectors should be related with nonlinear parameter and the
amplitude in a certain way. Then, RWs without any other type waves,
could be observed possibly  on the certain backgrounds. Meanwhile,
the ideal initial conditions for the RW are presented in the coupled
system. The parameter $\lambda$ which mainly determines the initial
shape of vector nonlinear waves, should be set with
\begin{equation}
\lambda=-\frac{k_1+k_2}{2} +I
\frac{3\sqrt{3}}{2} \sqrt{g_1} s_1.
\end{equation}

 With these
certain conditions, the generic form of vector RWs could be given as
\begin{eqnarray}
\psi_1&=&[1+\frac{3 \sqrt{3g_1}s_1 W_1(t,z)}{1+|u|^2+|v|^2}] s_1 e^{i \theta_1(t,z)} ,\\
\psi_2&=&[1+\frac{3 \sqrt{3g_1}s_1 W_2(t,z)}{1+|u|^2+|v|^2}] s_2
e^{i \theta_2(t,z)},
\end{eqnarray}
\begin{figure}[htb]
\centering
\subfigure[]{\includegraphics[height=30mm,width=40mm]{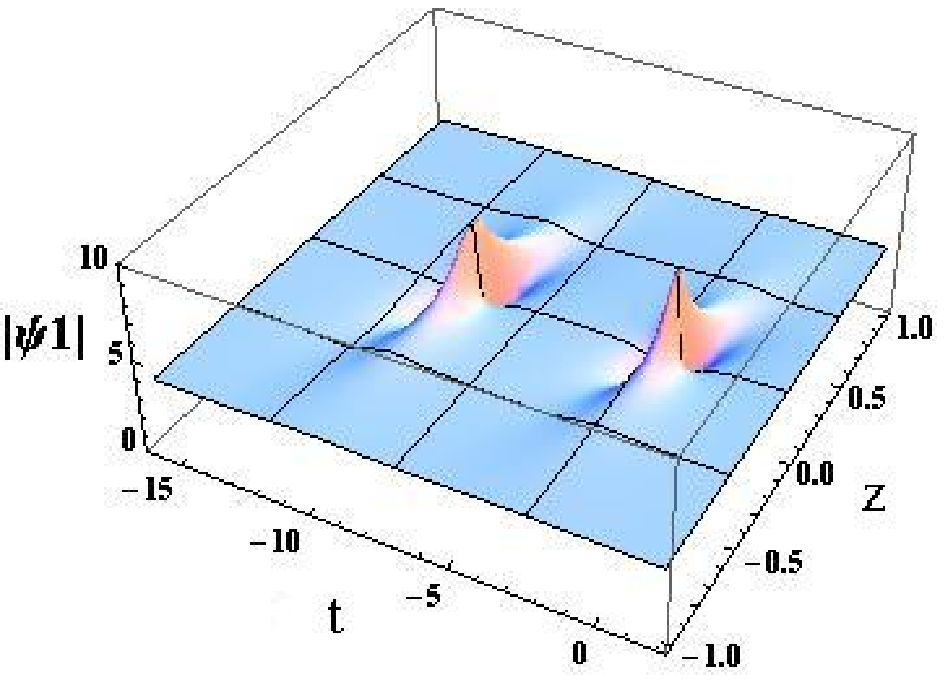}}
\hfil
\subfigure[]{\includegraphics[height=30mm,width=40mm]{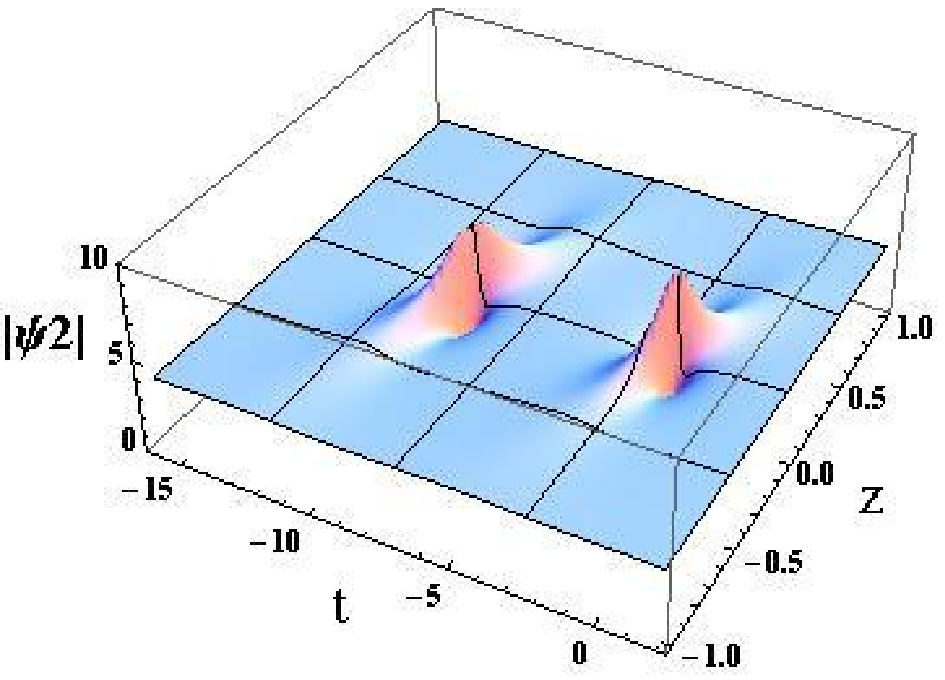}}
\caption{(color online)   The evolution plot of two RWs in coupled
system, (a) for two similar RWs in $\psi_1$ component, (b) for two
similar RWs in $\psi_2$ component. The coefficients are $ g_1 = g_2
= 0.25, s_1 =s_2=4, A_1 = 1, A_2 = 1, A_3 = 0.2, k_1 = 2.05$, and $
k_2 = 0.05$.}
\end{figure}

where $u=W_1(t,z) \sqrt{g_1} s_1$, and $v=W_2(t,z) \sqrt{g_2} s_2$.
The expressions of $W_{1,2}(t,z)$ are all rational functions,
presented in Appendix B. Therefore, the vector waves solution could
be vector RWs solution, which can be verified by the following RWs
plots. When $g_1=g_2=1$, the RWs solution could become the RWs in
\cite{Ling2}. When $A_3=0$, only one RW can be observed in both
components, shown in Fig. 5. The density of them are distinct in
temporal-spatial distribution, and their structures are similar to
the well-known RW in single-component system. When $A_3\neq0$, there
are two RWs appearing in the temporal-spatial distribution, shown in
Fig. 6. There are two RWs appearing in temporal-spatial
distribution, which is very distinct from the higher-order RW in
uncoupled system. In the uncoupled systems, it is nor possible to
observe just two RWs appearing in the whole temporal-spatial
distribution even for higher-order RWs \cite{Ling, J.K.Yang}.

When parameters $A_j$ vary, namely, the initial nonlinear waves are
slightly different, the RWs properties will be changed. For example,
when they are next to each other, the classical shape of RW
\cite{Akhmediev,Bludov}, called eye shape, could be changed, such as
fig. 7. However, the RW solution form is an ideal condition. It
could be very hard to realize them precisely. From the generic
solution Eq.(11) and (12), when the initial condition is a bit
different from the ideal initial condition, one can observe the
character of their evolution, such as Fig. 8. Comparing Fig. 6 and
Fig. 8, we can know that the vector RW could be seen as a limit case
of the general solution under the certain requirement condition.
When the initial conditions are similar to the ideal one, the
evolution of them will be close to the vector RWs. This is similar
to the case that RW is the limit of AB in the uncoupled system
\cite{Kibler}.

\begin{figure}[htb]
\centering
\subfigure[]{\includegraphics[height=30mm,width=40mm]{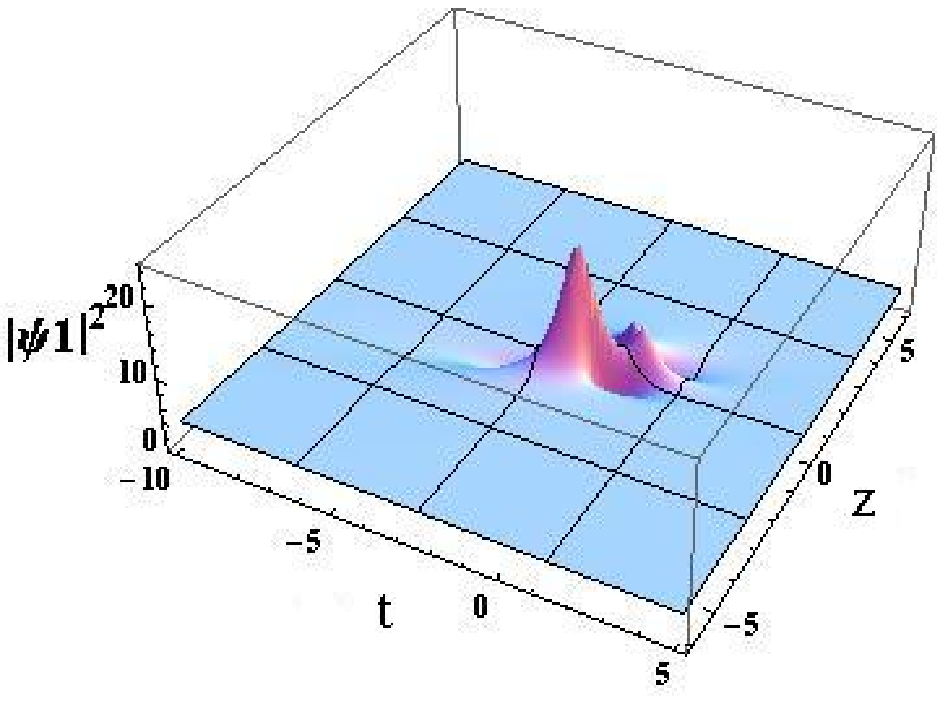}}
\hfil
\subfigure[]{\includegraphics[height=30mm,width=40mm]{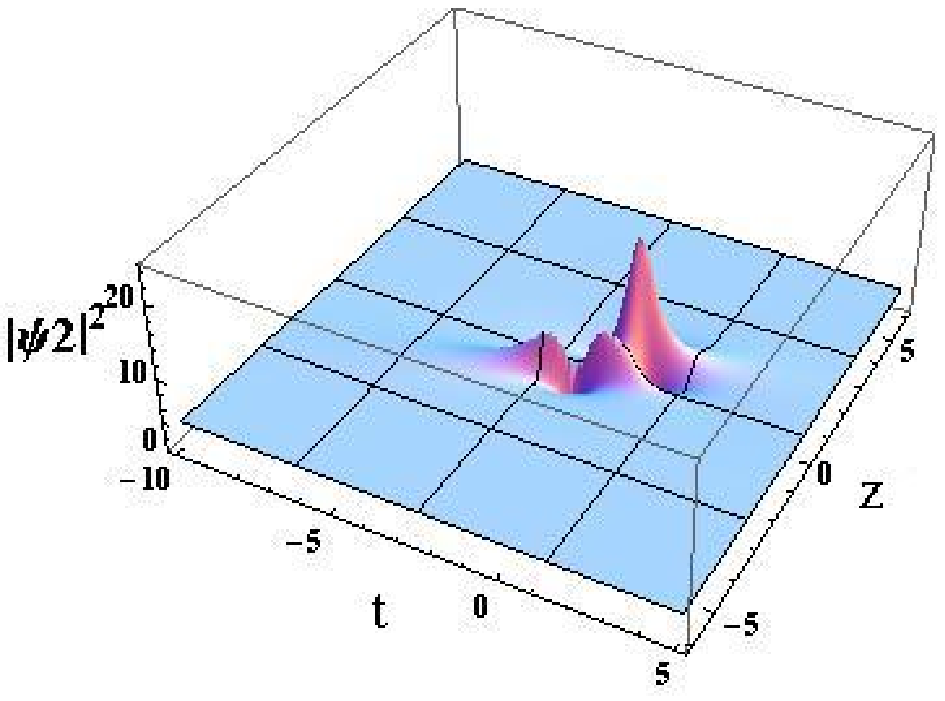}}
\caption{(color online)   The evolution plot of two vector RWs in
the coupled system, (a) for two different RWs in $\psi_1$ component,
(b) for two different RWs in $\psi_2$ component. It is seen that
dark RW could be observed in one component of the coupled system.
The coefficients are $ g_1 = g_2 = 0.5, s_1 = s_2 = 2, A_1 = 0, A_2
= 0, A_3 = 158, k_1 = 0.807$, and $ k_2 = 0.05$.}
\end{figure}

\begin{figure}[htb]
 \epsfxsize 45mm \epsfysize 30mm \epsffile{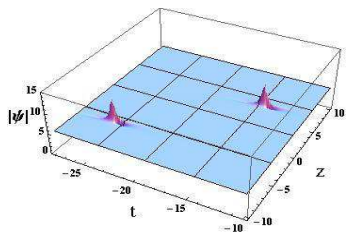}
 \\
\caption{(Color online) The evolution of whole ABs-ABs density in
the coupled system. The initial condition approaches the ideal
condition for RWs, and the corresponding required backgrounds are
set. It is seen that the evolution of them is close to RWs in Fig.6.
The explicit coefficients are $a_0 = -1.05, b_0 = 5.196, g_1 = g_2 =
0.25, s_1 =s_2=4, A_1 = 1, A_2 = 1, A_3 = 6, k_1 = 2.05$, and $ k_2
= 0.05$.}
\end{figure}

\section{ Possibilities to observe these vector nonlinear waves}

Considering the experiments on RW in nonlinear fibers with anomalous
GVD \cite{Kibler,Solli,Dudley}, which have shown that the simple
scalar NLS could describe nonlinear waves in nonlinear fibers well,
we expect that these different vector nonlinear waves could be
observed in two-mode nonlinear fibers. As an example, we discuss how
to observe the vector RWs. One can introduce two distinct modes to a
nonlinear fiber operating in the anomalous GVD regime
\cite{Afanasyev,Ueda}, then the ideal condition could be given in
detail. Firstly, we assume the nonlinear coefficients for the two
modes are equal, namely, $g_1=g_2=0.25$. The GVD coefficients are
$\sigma_1=\sigma_2=1$ in the normalized units. The spontaneous
development of RW seeded from some perturbation should be on the
continuous waves as the ones in \cite{Kibler,Solli,Dudley}.
Experimentally, one can use proper picosecond or nanosecond pulses
which are close to the ideal continuous waves for RW development
\cite{Dudley2}. The amplitudes of the backgrounds should be equal,
$s_1=s_2=1$ in the normalized units. One mode laser's frequency
$k_2=0.05$, and the other's is $k_1=k_2+0.5$. Then the ideal initial
optical shape can be given by Eq.(19) and (20) with
$\lambda=-\frac{k_1+k_2}{2} +I \frac{3\sqrt{3}}{4}$. Under these
conditions, the vector RWs could be observed in the nonlinear fiber.
The RW distribution is on the retarded time dimension, similar to
temporal optical solitons.

On the other hand, it is well known that there are spatial optical
solitons in planar waveguide. The similar conditions can be derived
directly through coordinates transformation for the coupled NLS in a
planar waveguide \cite{Cambournac}. The RWs in multi-mode planar
waveguide could be observed too. Additionally, considering the
studies on two-components Bose-Einstein condensate(BEC)
\cite{Doktorov,Das,Becker}, we expect that the vector RWs could also
be observed with the condition.

\section{Discussion and Conclusion}
  In summary, we obtain B-DAB, ABs-ABs, B-DRW, AB-ABRW, and vector RWs solutions for the coupled model.
  The corresponding conditions for their emergence are presented explicitly.
In particular, when the parameter $\lambda$, which mainly determines
the shape of initial signals, satisfies Eq.(13), there are ideal
initial signals existing which can evolve to one RW with other types
nonlinear waves, namely, B-DRW or AB-ABRW, etc. It provides
possibilities to observe interactions between RW and other type
nonlinear waves. This is an interesting subject for RW studies,
since these phenomena are impossible to be observed in the
single-component nonlinear systems.

Just under the conditions Eq.(16) and (17), which are the certain
requirements on both backgrounds of the two components, the initial
condition approaching the ideal one Eq.(19) and (20) with $\lambda$
given by Eq.(18), would evolve to the vector RWs with no other type
nonlinear waves. From the ideal condition, one can know that the two
vector RWs can not exist when the two backgrounds are relatively
static. This is very different from RW in the uncoupled system, for
which there is no requirement on nontrivial background
\cite{Kibler}. Additionally, dark RW, predicted in mixed BEC system
through numeric stimulation in \cite{Bludov2}, is observed
analytically here. Moreover, one possible way to observe vector RWs
in a two-mode nonlinear fiber is presented here.

Recently,
 the higher-order modulation instability with RWs has been observed
 in nonlinear fiber optics in \cite{Erkintalo}. There are many
 possibilities to observe similar phenomena in coupled nonlinear
 fiber system. We will continue to study this subject in the future work.

\section*{Appendix A: The analytic form for B-DRW AND AB-ABRW WAVES}
The RW with other type solitons could be derived as

\begin{eqnarray}
\psi_1&=&\psi_{10}-\frac{1}{\sqrt{g_1}}\frac{i(\lambda-\bar{\lambda})\Phi_1
\bar{\Phi}_2}{|\Phi_1|^2
+|\Phi_2|^2+|\Phi_3|^2},\\
\psi_2&=&\psi_{20}-\frac{1}{\sqrt{g_2}}\frac{i(\lambda-\bar{\lambda})\Phi_1\bar{\Phi}_3}{|\Phi_1|^2
+|\Phi_2|^2+|\Phi_3|^2},
\end{eqnarray}
where
\begin{eqnarray}
\Phi_1[t,z]&=&(\Phi_{01}+ \Phi_{02}+\Phi_{03})\nonumber\\
&&\times \text{Exp}\left[\frac{I}{3} (\theta_1[t,z]+\theta
_2[t,z])\right],\nonumber\\
\Phi_2[t,z]&=&(\frac{-\sqrt{g_1}s_1}{\tau_1-I \lambda /3 -I
(2k_1-k_2)/3} \Phi
_{01}[t,z]\nonumber\\
&&+\frac{-\sqrt{g_1}s_1}{\tau_2-I \lambda /3 -I (2k_1-k_2)/3}
\Phi_{02}[t,z]\nonumber\\
&&+\frac{-\sqrt{g_1}s_1-\frac{-\sqrt{g_1}s_1}{\tau_2-I \lambda /3 -I
(2k_1-k_2)/3}}{\tau_2
-I \lambda /3 -I (2k_1-k_2)/3} \Phi_{03}[t,z])\nonumber\\
&&\times \text{Exp}\left[\frac{I}{3}
(\theta_2[t,z]-2 \theta_1[t,z])\right],\nonumber\\
\Phi_3[t,z]&=&(\frac{-\sqrt{g_2}s_2}{\tau_1-I \lambda /3 -I
(2k_2-k_1)/3} \Phi
_{01}[t,z]\nonumber\\
&&+\frac{-\sqrt{g_2}s_2}{\tau_2-I \lambda /3 -I (2k_2-k_1)/3}
\Phi_{02}[t,z]\nonumber\\
&&+\frac{-\sqrt{g_2}s_2-\frac{-\sqrt{g_2}s_2}{\tau_2-I \lambda /3 -I
(2k_2-k_1)/3}}{\tau_2 -I \lambda /3 -I (2k_2-k_1)/3} \Phi_{03}[t,z])\nonumber\\
&&\times \text{Exp}\left[\frac{I}{3} (\theta_1[t,z]-2 \theta
_2[t,z])\right],\nonumber
 \end{eqnarray}
and
\begin{eqnarray}
\Phi_{01}&=&A_1 \text{Exp}[\tau_1 t + I \tau_1^2 z + 2 (\lambda - k_1 - k_2) \tau_1 z/3],\nonumber\\
\Phi_{02}&=&(A_3 t +2 A_3 I \tau_2 z +2/3  A_3(\lambda - k_1 - k_2)z
+A_2) \nonumber\\
&&\times \text{Exp}[\tau_2 t + I \tau_2^2 z + 2 (\lambda - k_1 - k_2) \tau_2 z/3], \nonumber\\
\Phi_{03}&=&
 A_3\text{Exp}[\tau_2 t + I \tau_2^2 z +
    2 (\lambda - k_1 - k_2) \tau_2 z/3].\nonumber
\end{eqnarray}
Between the above expressions, the parameter $\lambda$ is the
solution of Eq.(13).

\section*{Appendix B: The analytic form for $W_{1,2}$}
The expressions of $W_{1,2}$ are
\begin{eqnarray}
W_{1,2}(t,z)&=&\frac{1}{K_{1,2}}+\frac{1}{K_{1,2}^2}+\frac{P_{1,2}(t,z)}{Q(t,z)}\nonumber
\end{eqnarray}
where \begin{eqnarray} K_1&=&\frac{i(k_1-k_2)}{2}-\frac{\sqrt{3
g_1}s_1}{2},\nonumber\\
K_2&=&\frac{i(k_2-k_1)}{2}-\frac{\sqrt{3 g_1} s_1}{2},\nonumber
\end{eqnarray}
 and
\begin{eqnarray}
Q(t,z)&=& \frac{1}{2} A_3 t^2+\frac{2}{9} (\lambda-k_1-k_2)^2A_3
z^2+(A_2+A_3) t\nonumber\\
&&+\frac{2}{3}
(\lambda-k_1-k_2) (A_2+A_3) z+i A_3 z\nonumber\\
&&+\frac{2}{3} (\lambda-k_1-k_2) A_3 t z+A_1+A_2+A_3,\nonumber\\
P_{1,2}(t,z)&=&\frac{A_3}{K_{1,2}^3}-\frac{1}{K_{1,2}^2}
M(t,z),\nonumber\\
M(t,z)&=&\frac{1}{2} A_3 t^2+\frac{2}{9} (\lambda-k_1-k_2)^2A_3
z^2+A_2 t\nonumber\\
&&+\frac{2}{3}
(\lambda-k_1-k_2) A_2 z+i A_3 z\nonumber\\
&&+\frac{2}{3} (\lambda-k_1-k_2) A_3 t z+A_1.\nonumber
\end{eqnarray}
Between the above expressions, the parameter $\lambda$ is given by
Eq.(18).

\section*{Acknowledgments}
 We are grateful to Professor L.B. Fu and Y.J. Chen for helping in theoretical analysis. This work
is supported by the National Fundamental Research Program of China
(Contact No. 2011CB921503), the National Science Foundation of China
(Contact Nos. 11274051, 91021021).

\end{document}